# Multicore Dynamic Kernel Modules Attachment Technique for Kernel Performance Enhancement

Mohamed Farag

Department of Computer Science, Maharishi University of Management, USA
mfarrag@freebsd.org

## *ABSTRACT*

*Traditional monolithic kernels dominated kernel structures for long time along with small sized kernels, few hardware companies and limited kernel functionalities. Monolithic kernel structure was not applicable when the number of hardware companies increased and kernel services consumed by different users for many purposes. One of the biggest disadvantages of the monolithic kernels is the inflexibility due to the need to include all the available modules in kernel compilation causing high time consuming. Lately, new kernel structure was introduced through multicore operating systems. Unfortunately, many multicore operating systems such as barrelfish and FOS are experimental. This paper aims to simulate the performance of multicore hybrid kernels through dynamic kernel module customized attachment/ deattachment for multicore machines. In addition, this paper proposes a new technique for loading dynamic kernel modules based on the user needs and machine capabilities.*

## *KEYWORDS*

*Multicore, Kernel, Dynamic Module, UNIX, Linux.*

## 1. INTRODUCTION

For long time, monolithic kernels dominated the operating system structures along with the era of small sized kernels, limited hardware companies and limited kernel functionalities. Due to the huge number of kernel modules developed to the operating system, it was a good idea to extract regions of the kernel and attach them on-demand for minimizing the space and the number of handled modules for achieving greater performance. So, the idea of dynamic kernel modules was a great one to encapsulate some kernel regions into "pluggable" kernel modules. Dynamic kernel module is an object file that contains code to extend the running kernel, or so-called base kernel, of an operating system. Dynamic kernel modules are typically used to add support for new hardware and/or filesystems, or for adding system calls. Most current Unix-like systems, and Microsoft Windows, support loadable kernel modules, although they might use a different name for them, such as kernel loadable module (kld) in FreeBSD, kernel extension (kext) in OS X and Loadable Kernel Modules (LKM) in Linux. They are also known as Kernel Loadable Modules or KLM [1]. This ability also helps software developers to develop new parts of the kernel without constantly rebooting to test their changes [2]. Dynamic Kernel Module is a great deal for unnecessary drivers for kernel initialization such as network card drivers and sound drivers. The basic structure of current dynamic kernel modules includes loader handler, module declaration and optional finalization function. The loader handler loads module to the kernel, module declaration contains the code implementing module functionality and finalization function performs all the module de-allocation process for freeing memory. Lately, kernels support multi-core feature or so-called SMP kernels*. However, SMP kernels have some issues because if one CPU fails, the entire SMP system is down.

## 2. PROBLEM BEHAVIOR

Nowadays, there is numerous number of kernel modules causing high time consuming to load all of them at once. Some users might ask to load modules not supported by hardware and unfortunately, most of the systems would load those modules and don't prompt users for that. In addition, there's excessive overhead to load all the modules on one core while it has multi-core kernel support. Another communication overhead source would be loading some kernel modules on one core without handling dependencies between kernel modules. What about if a kernel module which has descendants failed?! What about if a program is dynamically divided and executed on multiple CPU cores in SMP kernels? It is impossible to predict the execution order of individual processes or which CPU core executes what process so real-time behavior may not be guaranteed for some processes. Dynamic kernel modules are still working in manual manner and without communication organization or fault tolerance cover. This traditional manner leads us to search for more efficient alternatives. In the next section, a generic multi-stage solution for handling the attachment of kernel modules automatically with respect to correctness and efficiency will be discussed.

## 3. SUGGESTED MULTI-STAGE SOLUTION

In this multi-stage solution, we will attach the kernel modules based on "physical hardware existence" and available memory space with respect to the mandatory kernel module for basic operation. First, we will start handling the problem in single core environment and after that we will solve it in multicore environment. Two algorithms will be introduced; first one is used for indexing (i.e. specifying which dynamic kernel modules will be loaded) and the second one specifies how dynamic modules will be loaded automatically. Some issues are solved in these algorithms such as module dependencies across single/many core(s), retrieval of hardware information and inefficient loading of kernel modules. Some issues won't be discussed because only generic issues are discussed and other issues are platform dependant such as retrieving kernel modules from the corresponding directories and adding module to specific core. In that case, minor changes are considered to the SMP kernel to take the control of process distribution across cores. Notice that for most of operating systems there's a macro used to attach the dependencies at module creation; for instance, MODULE_DEPEND macro is used to handle module dependencies in FreeBSD (UNIX-Like System) [2]. On the other hand, some tools are used to fetch hardware information from BIOS such as lshw in Linux System and dmesg or syscons in FreeBSD and UNIX-Like Systems [3]. In most systems, kernel modules can be found in one directory or two at most so that saves lots of efforts to track all the available kernel modules. The location of the kernel modules depends on the operating system. For example, Kernel modules are located in /boot/kernel or /boot/modules in FreeBSD and most of UNIX-Like systems and located in pcmcia/ and kernel/ directories in Linux systems [4]. The state of the kernel module (loaded/unloaded) affects the memory size used by the kernel. This technique limits insufficient kernel modules through comparing all kernel modules with the one stored in the BIOS. In addition, this paper discusses some kernel regions can't be loaded at run-time and are required to be part of the base-kernel. In general, all the stages have the same structure so they have module registration technique used to register loaded modules, module loading technique which is used to load required modules and index file containing the flags of the modules to be loaded in the next boot time. This procedure starts with index file creation using module registration. After that, modules will be loaded from index file each time kernel initialized using module loading technique. Next we will consider stage-0 solution for dynamic kernel modules attachment.

## 1.1. Stage-0 Solution

This technique is adapted for both UNIX-Like and Linux systems. In this solution, the correctness is the only matter we care about since it's the first goal seeked in solving all the problems. The algorithm used for loading dynamic kernel modules is described in the following steps using depth first approach assuming the file is already filled before:

1. Load all kernel modules alphabetically from the module directories neglecting module names ending with .symbols in it because these files are just helper files for the original modules
2. Map each index in the index file to loadable kernel module. In this step we will store the kernel module names -eliminating the extension- in a proper data structure –arrays and maps are good candidate- (ideally, each possible kernel module should have a bit indicating whether it's loaded or unloaded).
3. Read device names from BIOS using hardware information retrieval tool (lshw, syscons, dmesg, sysctl or sysinfo) and save them into proper data structure such as array.
4. Read module name from kernel module array
5. Loop through the kernel modules array
6. Check flag status in the index file (raised/not raised).
7. If flag is raised
8. match = check_hardware_support()
9. If not match
10. Continue to the next iteration
11. End if
12. handle_module()
13. End if
14. Read next module name from kernel module array
15. End loop

**check_hardware_support()**

1. Use the module name stored as array value in kernel
   modules array
2. Loop through the hardware information array
3. if there's a match between the value and
corresponding one stored in the hardware info array.
4. return true
5. End if
6. End loop
7. return false

**handle_module()**

1. Grab module dependencies using lsmod and modinfo tools.
2. Loop for module dependencies
3. Call handle_module()
4. Load module
5. End loop

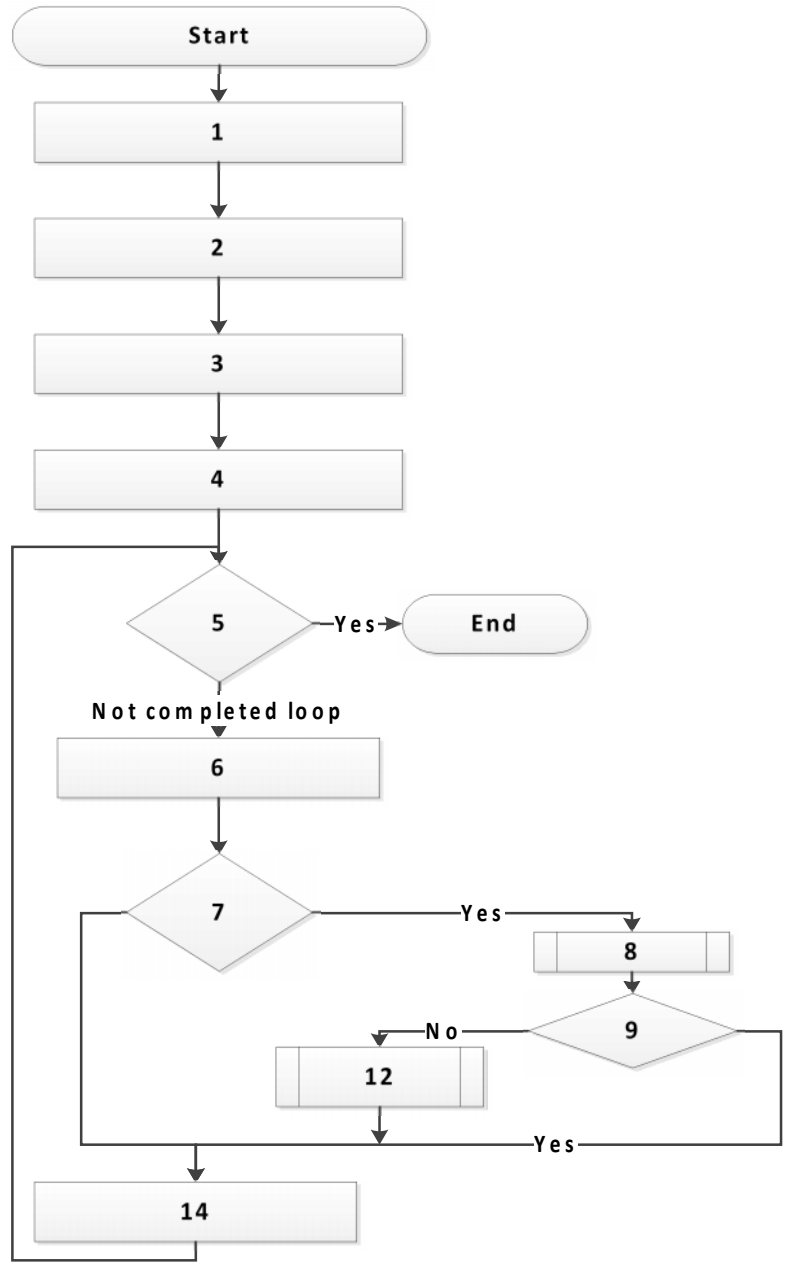

Fig 1.0 Module Loading at Stage-0 Solution

### 3.1.1. Restrictions

• The kernel module registration technique should run after each change to the kernel modules directory.
• This technique assumes no direct/indirect circular dependencies.

The selection of the hardware information data structure holding values retrieved from BIOS depends on the tool used for grabbing BIOS information; some of them can be stored directly in simple data structures such as sysctl hw but others need some manipulation. However, this is not the end! Next, we need to clarify the mystery of the index file created by module registration technique. However, it's simple one since complexity was moved to module loading portion so module registration will be trivial one:

1. Load Kernel modules directory into an array
2. Loop for the array
3. Ask user to load/unload
4. If user selects load
5. Flag the position of module by 1 in the file
6. Else
7. Flag the position of module by 0 in the file
8. End If
9. End Loop

### 3.1.2. Stage-0 Solution Drawbacks

The biggest disadvantage of this procedure is the performance which can be enhanced through simple rephrasing of the program structure. In addition, module registration is working in very trivial manner.

This solution is proposed to enhance the performance of stage-0 solution on single core machines. Stage-1 solution moves the complexity to module registration technique instead of module loading technique since it's slightly more often to use module loading more than module registration. In this solution, each module will occupy at least one byte to represent the depth of module dependencies. The main idea of this solution is to determine the most independent modules and assign them higher priorities than other modules. Level 1 is higher priority than level 2 and so on. In module loading process, the independent module will be loaded first and this guarantees no problems with module dependencies. The "level structure" is important for systems that don't have implicit module dependency detection. Module registration technique is discussed in the following steps:

1. Load kernel modules from module directories into a global hashmap with zero associated values //use malloc
2. Read first module from map.
3. Loop for modules in the map
4. Initialize count with zero //module not supported by hardware
5. Ask user to load/unload // trivial selection approach
6. If load // if user selected to load it
7. match = check_hardware_support() // flag variable

8. if match   // if this module is supported by hardware
9. reinitialize count with 0
10. globalLevel = handle_dependancy(module_name, count)
11. store globalLevel as a value for key module_name in the map
12. end if
13. end if
14. read next module from the map
15. end loop
16. write map to index file

int handle_dependancy(module_name, count)
1. detect module dependency // lsmod & modinfo
2. store new modules in new_array
3. value = count + 1
4. read first module from new_array into new_module
5. loop through new_array
6. load value corresponding to module name from the map
level = handle_dependancy ( new_module, newCount)
7. if (level > value) //make sure it's the lowest priority
8. store level + 1 as value for key module_name in the map
9. return level
10. end if
11. read next module from new_array into new_module
End Loop
12. return value //returned if there's no module dependency for this module
13. End function

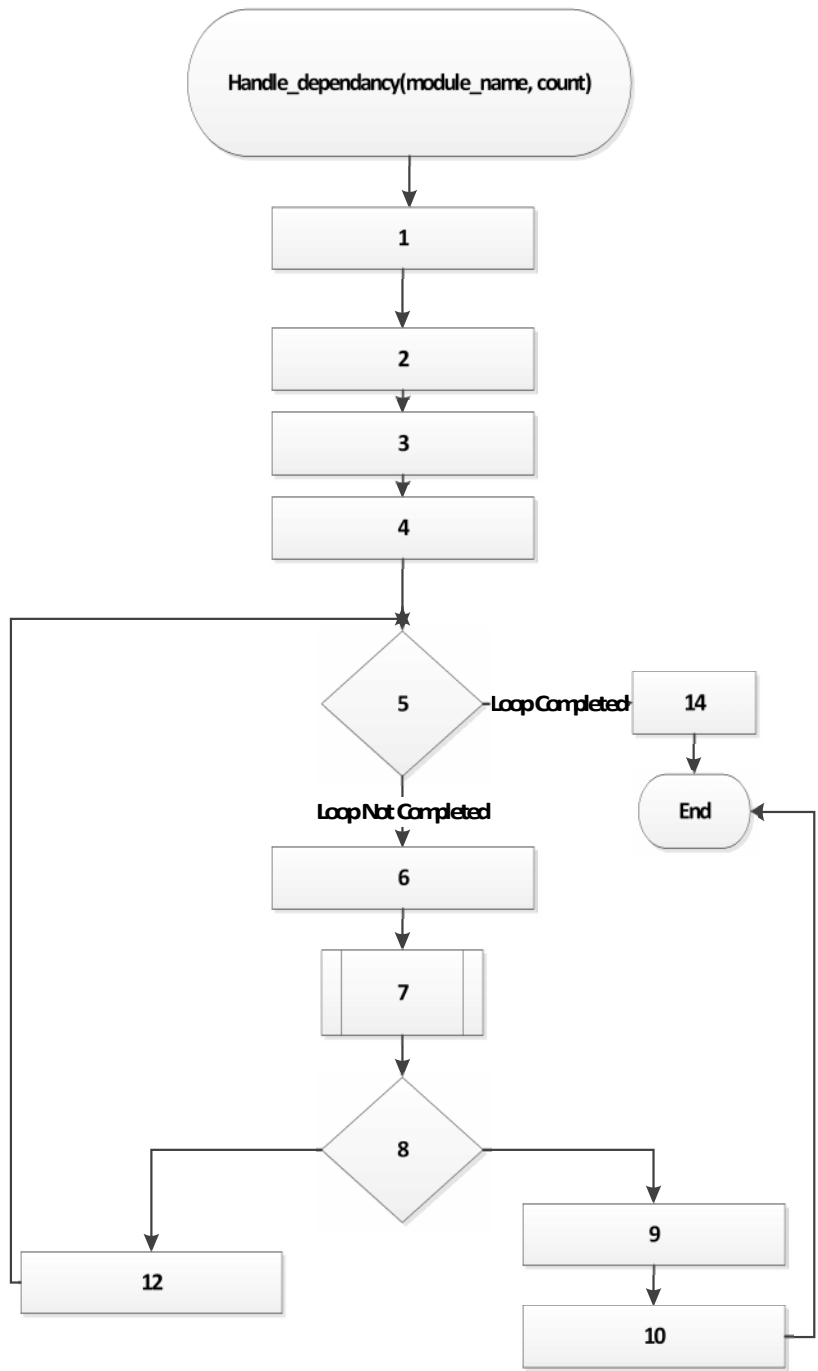

Fig 2.0 handle_dependancy in Module Registration

Notice that count variable is used to represent the depth of module dependencies i.e. differentiate layers of dependencies. Count solves the module dependencies through associating the most independent modules with lower numbers and the most dependant modules with relatively larger numbers. In addition, it is initialized by 1 for modules matching the physical hardware. Otherwise, it remains zero. The usage of count variable facilitates the differentiation between independent modules and incorrect modules.

| Count Variable | |
|---|---|
| 0 | Not supported by hardware |
| 1 | Independent module |
| 2 – 255 | Dependant module |

Table 1.0 count variable possible values

On the other hand, module loading technique seems to be much simpler than the one was suggested in Stage-0 solution. The following steps describe the operation of module loading technique:

1. load kernel modules from kernel directory into modules array
2. load values from index file into values array
3. initialize depth by 1
4. do { //start outer loop
5. initialize loop_counter by 0
6. initialize flag with false //used to check if reached the end limit or not
7. loop through modules array
8. flag = true //at least entered the loop once
9. if depth equals to values[loop_counter]
10. load modules[loop_counter] //kldload
11. end if
12. increment loop_counter by 1
13. end loop
14. increment depth by 1
15. } while (flag != false)  //end of outer loop

### 3.2.1. Advantages of Stage-1 Solution

For typical scenario including one time of module registering and four times of module loading, the overall performance is improved by around 150% because of the efficient allocation of kernel modules in addition to usage of simpler module loading technique much more than the complicated module registration. However, previous techniques achieve good results for single core machines but they are outdated now.

### 3.3. Stage-2 Solution

There's noticeable growth in the number of cores in machines so, multi-core technique for automatic attachment of dynamic kernel modules will be introduced. Unfortunately, Stage-1

International Journal of Computer Science & Information Technology (IJCSIT) Vol 4, No 4, August 2012solution can't be parallelized because it has loop carried dependencies caused by depth variable. The following steps describe the modified Stage-0 technique for multicore architectures. First, let's have a look in module registration technique (No changes with Stage-0 solution):

1. Load Kernel modules directory into an array
2. read module from the array
3. Loop through the array
4. Ask user to load/unload
5. If user selects load
6. Flag the position of module by 1 in index file
7. Else
8. Flag the position of module by 0 in index file
9. End if
10. read next module from the array
11. End Loop

This one is pretty similar to stage-0 solution one because there is no opportunity for parallelization because module registration depends on user response. The next step is to discuss the module loading in the following steps:

1. Create pthread for each core (Assign one for each core)
2. Assign one thread for the following
a. Load all kernel modules alphabetically from the module directories neglecting module names ending with .symbols in it because these files are just helper files for the original modules.
b. Map each index in the index file to loadable kernel module. In this step we will store the kernel module names -eliminating the extension- in a proper data structure –arrays and maps are good candidate- (ideally, each possible kernel module should have a bit indicating whether it's loaded or unloaded).
3. Assign another thread to read device names using hardware information retrieval tool (lshw, syscons, dmesg, sysctl or sysinfo) and save them into proper data structure.
4. Wait for both threads to complete.
5. In each thread, Loop through kernel modules array
6. Read module name from it
7. Check flag status in the index file (raised/not raised).
8. If flag is raised
9. match = check_hardware_support()
10. If not match
11. Continue to the next iteration
12. End if
13. Lock
14. handle_module()
15. Unlock
16. End if
17. End loop

Note: check_hardware_support() and handle_module() modules are similar to the ones implemented in stage-0.

55

International Journal of Computer Science & Information Technology (IJCSIT) Vol 4, No 4, August 2012

Locking handle_module() subroutine leads to huge performance degradation. This leads us to present locking-free algorithm in Stage-3 solution.

## 3.4. Stage-3 Solution

The main disadvantage with stage-2 solution is the locking mechanism used to lock modules so, free-locking module loading technique is described in the following steps:

1. Check number of modules in the kernel modules directory
2. Calculate step to be number of kernel modules divided by number of cores - 1
3. Create array of pthreads with size of number of cores
4. Assign all threads except one to map each index in the kernel modules loading file to loadable kernel module in the directory. This process includes two steps:
a. Load all kernel modules alphabetically from the module directories neglecting module names ending with .symbols in it because these files are just helper files for the original modules.
b. Map each index in the index file to loadable kernel module. In this step we will store the kernel module names -eliminating the extension- in a proper data structure –arrays and maps are good candidate- (ideally, each possible kernel module should have a bit indicating whether it's loaded or unloaded).
5. Assign last thread to read hardware names using hardware information retrieval tool (lshw, syscons, dmesg, sysctl or sysinfo) and save them into proper data structure.
6. Wait for all threads to finish.
7. Calculate start point and end point for each thread.
start_point = thread_index*step
end_point = (thread_index+1) * step
8. Assign (Total number of threads -1) to the following array
9. In each thread, loop through the kernel modules array from start_point to end_point
10. Read module name from it
11. Check if flag is raised in the flags holder file.
12. If flag is raised
13. match = check_hardware_support()
14. If not  match
15. Continue to the next iteration
16. End if
17. handle_module()
18. End if
19. End loop

    **handle_module()**

    1. if module is loaded //make sure it's not loaded by another thread
    2. return;
    3. Detect module dependencies
    4. Loop for module dependencies
    5. Call handle_module()
    6. If module is not loaded
    7. Load module

8. End If
9. End loop

Note: check_hardware_support() module is similar to the one implemented in stage-0.
We should clarify two things in the last algorithm. The main difference between stage-3 and stage-2 is locking. Stage-3 Algorithm doesn't require any locking because it assigns specific range of modules to each thread. In addition, we check these modules not being loaded by other threads as module dependencies. Module loading checking occurs during the execution of handle_module() function and we don't need to use any locks to check if module is loaded because it's read operation.

### 3.4.1. Stage-3 Restrictions

From practical experiments, this algorithm is efficient for machines with 4 cores at least. Module registration is the same as traditional ones so we don't need to rewrite it. Stage 3 solution provides a lot of performance-wise benefits through avoiding collisions, locks and duplications. This solution offers a great speedup over the sequential version.

## 4. SPACE SAVING CALCULATIONS

Here's an example showing sample case for space saving through eliminating some of the unneeded modules. FreeBSD 8.0 Machine was used to calculate some space savings using our technique. This technique can be applied at the very beginning of the system to achieve great performance and huge savings. For example, we could decrement the kernel by 2331 KB when we specified specific machine architecture. Another 2112 KB could be deducted from kernel size after unloading INET6 kernel module. However, in FreeBSD, some kernel regions can't be isolated from the kernel such as FFS, GEOM_GPT for handling disk partitioning and root account related modules.

## 5. PERFORMANCE MEASUREMENTS

The multi-stage solution was tested using loggers. In single core machines, it's easy to log the system time to files by adding few lines to the loader handler function. The following code snippet describes logging execution time in addition to counting number of modules loaded by the system:

```
static int skel_loader(struct module *m, int what, void *arg)
{
  If (counter == 0) { //first loaded module
gettimeofday(&time_now,NULL);
append_to_file(&time_now); //write the start time of the system loading
}
else if (counter == TOTAL_NUMBER_OF_MODULES) { // check if the counter reaches
    the total number of modules should be loaded.
gettimeofday(&time_now,NULL);
append_to_file(&time_now); //write the end time of the system loading
else {
counter = counter + 1;
}
// rest of the code
..
}
```

Note: TOTAL_NUMBER_OF_MODULES is used for test purposes only. In real scenario, we don't know how many modules should be loaded for the fact that some modules are rejected due to failing to find hardware support. Now, the counter variable definition is missing so in order to avoid the issues of reinitializing variables several times, we will initialize it as global variable in the scheduler initialization function which is guaranteed to be executed first and only once.

In multicore machines, it's slightly more difficult to initialize the counter variable because the code is distributed among several cores and each has its own copy in some machines. Message Passing Interface "MPI" is used to solve variable reinitialization to guarantee correct behavior and that was excluded from our calculations. The following statistics show the average time needed to execute each technique after normalization process.

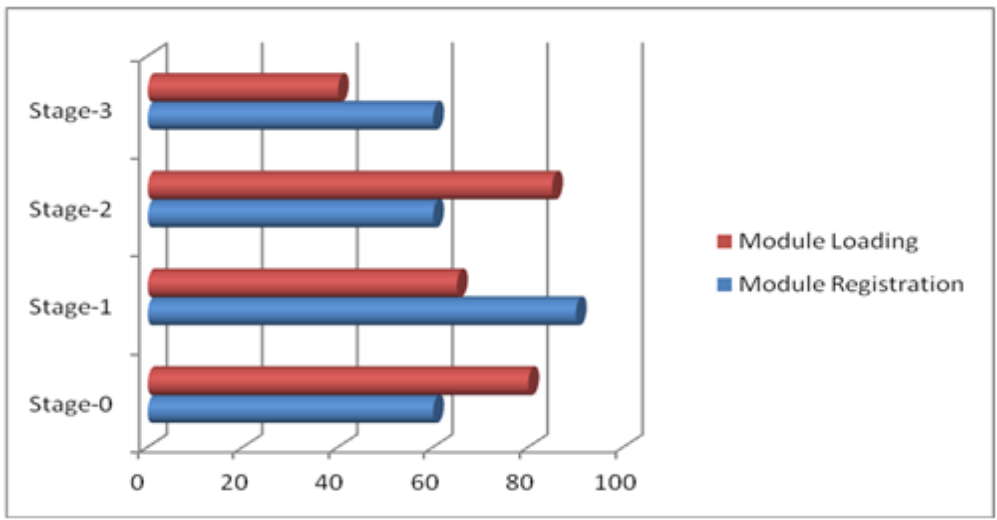

Fig 3.0 Normalized Performance Measurement on FreeBSD 8.0, i7 with hyper-threading support

## 6. RELATED WORK

Dynamic Kernel Modules Support Project (DKMS) produced by Dell used for the same purposes but the technique we proposed solves issues with handling modules on multicore machine and have complete different view in module customization [3]. Also, it differs from other asynchronous parallel initialization systems that don't handle module customization and BIOS checking. In addition, this technique offers much more than what offered by MODULE_DEPEND

## 7. Future Work

The techniques proposed in the paper are just the beginning for parallel kernel management system leading us to multicore kernel management unit. In addition, the module registration based on user response will be enhanced to use predefined configuration files or extract the configuration from compiled kernels.

## 8. CONCLUSION

This paper proposes one of the kernel management techniques toward minimizing the kernel size and achieving greater performance. It discusses the dynamic kernel modules attachment process in smooth and well-defined process moving in step-by-step manner from the single core to many core machines. These techniques have been proved practically to be large step for promising multicore kernel management unit.

## 9. ACKNOWLEDGMENT

I thank Warner Losh, FreeBSD Core Team Member and Senior Software Engineer for his efforts and great support for me through mentoring me in Google Summer of Code 2010.

**Author**

"**Mohamed Farag** is post graduate student in Maharishi University of Management in USA. In the first six months of 2012, Mohamed worked as teaching assistant in Maharishi University of Management. He also worked as instructor in Ain Shams University in Egypt during the year 2011.  In 2010, Mohamed received Google Summer of Code award and was honored by the scientific community in Menoufia University in Egypt for his great contributions. In addition, Mohamed received "The Best Programming Project" award in Egyptian Universities Summit in 2010 and the same award in 2011. Mohamed has been an active contributor in FreeBSD community since May, 2010 and has led ArabBSD project since June, 2011. In 2012, Mohamed was selected to join the Institute for Computer Sciences, Social Informatics and Telecommunications Engineering (ICST), and International Association of Computer Science and Information Technology (IACSIT)."